\documentclass[11pt,a4paper]{article}
\usepackage[latin1]{inputenc}
\usepackage{amsmath}
\usepackage{amsfonts}
\usepackage{amssymb}
\usepackage{graphicx}
\usepackage{subfloat}
\DeclareGraphicsExtensions{.pdf}
\author{Sumit Ghosh}
\title{RN ADS}
\begin{document}

\title{{\bf{New type of phase transition in Reissner Nordstr{\"o}m - AdS black hole\\and its thermodynamic geometry }}}

\author{{\bf {\normalsize Rabin Banerjee}$
$\thanks{E-mail: rabin@bose.res.in}}, \,
{\bf {\normalsize  Sumit Ghosh}$
$\thanks{E-mail: sumit.ghosh@bose.res.in},\,
{\bf {\normalsize Dibakar Roychowdhury}$
$\thanks{E-mail: dibakar@bose.res.in}}}\\
{\normalsize S. N. Bose National Centre for Basic Sciences,}
\\{\normalsize JD Block, Sector III, Salt Lake, Kolkata-700098, India}
\\[0.3cm]} 

\maketitle
\begin{abstract}
The phase transition of a RN-AdS black hole is studied in details using Ehrenfest's equations. There is no discontinuity in entropy which signals a lack of any first order phase transition. We then show that although Ehrenfest's first equation is satisfied, the second is not, so that a true second order phase transition is also ruled out. However this deviation from the second Ehrenfest's equation, for a certain range of the black hole charge, indicates the existence of a new glassy type transition. We finally study the thermodynamic behaviour using state space geometry and find that the scalar curvature diverges exactly at those points where the heat capacity diverges. 
\end{abstract}
\section{\bf Introduction}

Black holes are the most exotic objects in physics. Even more surprising is their connection with thermodynamics. The identification of black holes as thermodynamic objects with physical temperature and entropy opened a  gate to a new realm. Once black holes are identified as thermodynamical objects, it is very natural to ask whether they also behave as familiar thermodynamic systems. A fascinating topic is the study of phase transition in black holes. Occurrence of a phase transition in black hole thermodynamics was observed long ago \cite{Davies,Pavon}. The phenomenon becomes more vivid when we study black holes in AdS space \cite{DPage,Peca}.

In usual thermodynamics, the various entities in a particular phase behave smoothly while they manifest a sharp discontinuity at the transition point. There are standard methods to study such phenomena which are based on the Clausius-Clapeyron scheme (first order phase transition) or the Ehrenfest's scheme (second order phase transition). In spite of several papers \cite{Davies}-\cite{Peca} on black hole phase transition, a systamatic study based on the afore stated schemes is lacking. Work in this direction was only commenced very recently \cite{Modak1,Modak2} 

An alternative way to study thermodynamics is using geometry. In 1979 George Ruppeiner \cite{Rupp} proposed a geometrical way to study  thermodynamics and statistical mechanics. Different aspects of thermodynamics and statistical mechanics of the system are encoded in the metric of the state space. Consequently various thermodynamic properties of the system can be obtained from the properties of this metric and curvature. It is successful in describing the behaviour of classical systems \cite{Rupp,Mijat} and recently it has been found to play an important role in explaining black hole thermodynamics \cite{Modak2},\cite{Mirza}-\cite{Rupp2}.

In this paper we have made a systematic study of the phase transition in RN-AdS black holes using Ehrenfest's scheme. We find a couple of points, dictated by the charge of the black hole, where the heat capacity diverges thereby indicating the presence of a phase transition. Absence of any discontinuity in entropy-temperature relationship eliminates the presence of any first order transition. We then perform a detailed analysis of the two Ehrenfest's equations using both analytical and graphical techniques. While the first Ehrenfest's equation is satisfied the second one is violated. Consequently this rules out a conventional second order phase transition. The deviation from the second Ehrenfest's relation is quantified by calculating the Prigogine-Defay (PD) ratio \cite{Gupta}. Interestingly we find that this ratio, for a certain range of the black hole charge, lies within the bound conforming to a glassy phase transition \cite{Jack,Th,Thu}.

The next part of our paper consists in analysing the above issues using the state space geometry. This approach was used earlier \cite{Aman}-\cite{Rupp2} in the context of black holes. Specifically, a divergence in the curvature scalar was associated with the presence of a phase transition. However the particular nature of the phase transition could not be identified. We have succeeded here. Also, as will be elaborated further this is neither a first nor a second order transition. It is a smeared second order transition called glassy phase transition. We find the curvature scalar diverges precisely at the same points where the heat capacity also diverges. We also observe that the expression for the curvature scalar suggests the existence of other poles where the curvature blows up but the heat capacity does not. We successfully show that the divergence of the curvature scalar at these points is physically disallowed by the nonextremality condition.

The organisation of the paper is as follows. Section 2 deals with the thermodynamics of RN-AdS black hole. In section 3 we have analysed the nature of the phase transition using Ehrenfest's scheme.    In section 4 we have studied the thermodynamics using state space geometry. Section 5 contains a discussion of our results. There are three appendices where certain technical details have been given. 
\section{Thermodynamics of Reissner-Nordstr{\"o}m black hole in AdS space}

Reissner-Nordstr{\"o}m black holes are charecterised by their mass (M) and charge (Q). The solution of RN-AdS black hole with a negative cosmological constant $ (\Lambda=-3/l^2) $ is defined by the line element,
\begin{equation}
ds^2 = \chi dt^2-\chi ^{-1}dr^2-r^2 \left(d \theta ^2 + \sin ^2 \theta d \phi^2 \right)
\end{equation} 
where
\begin{equation}
\chi = 1-\frac{2M}{r}+\frac{Q^2}{r^2}+\frac{r^2}{l^2}
\end{equation}
The asymptotic form of this line element is AdS. The outer horizon is defined by $\chi\left(r_+\right)=0$ which leads to the following expression for the mass of the black hole,
\begin{equation}
2M = r_+ + \frac{r_+ ^3}{l^2} + \frac{Q^2}{r_+}
\label{Mr}
\end{equation}
Using the semiclassical result for entropy , i.e.
\begin{equation}
S= \pi r_+^2
\label{S}
\end{equation} 
the expression for the black hole mass ({\ref{Mr}}) becomes
\begin{equation}
2M = \sqrt{\frac{S}{\pi}} + Q^2\sqrt{\frac{\pi}{S}} + \frac{1}{l^2}\left(\frac{S}{\pi}\right)^{\frac{3}{2}}
\label{M}
\end{equation}
It is now possible to determine the other thermodynamic entities using the basic thermodynamic relations
\begin{equation}
\delta M = T \delta S + \Phi \delta Q  \label{thermo} \\
\end{equation}
These are defined as
\begin{eqnarray}
T &= \left(\frac{\partial M}{\partial S}\right)_Q &= \frac{1}{4 \pi ^{3/2} S^{3/2}} \left(-\pi ^2 Q^2 +\pi S + 3S^2 / l^2 \right) \nonumber \\ \label{DEF1} 
\Phi &= \left(\frac{\partial M}{\partial Q}\right)_S &= Q \sqrt{\frac{\pi}{S}} \\
C_\Phi &= T\left(\frac{\partial S}{\partial T}\right)_\Phi &= 2S \frac{-Q^2 \pi ^2 + \pi S +3S^2/l^2}{Q^2 \pi ^2 - \pi S +3S^2/l^2} \nonumber 
\end{eqnarray}
where $ \Phi $ is the potential difference between the horizon and infinity, $T$ is the Hawking temperature, $S$ is the entropy and $C_\Phi$ is the heat capacity at constant potential of the black hole. 

For the convenience of further calculation we will scale the variables M,S,Q,T and $C_\Phi$ using the characteristic length $l$. We introduce a new set of variables defined as,
\begin{eqnarray}
m = \frac{M}{l} \hspace{1cm} q=\frac{Q}{l} \hspace{1cm} s=\frac{S}{l^2} \hspace{1cm} t=T l \hspace{1cm} c_\Phi = \frac{C_\Phi}{l^2}
\end{eqnarray} 
and we will call these new variables as the mass (m), charge (q), entropy (s)  temperature (t) and heat capacity ($c_\Phi$) of the black hole.

In terms of these new variables (\ref{M}) and ({\ref{DEF1}}) take the forms,
\begin{eqnarray}
2m &=& \sqrt{\frac{s}{\pi}} + q^2\sqrt{\frac{\pi}{s}} + \left(\frac{s}{\pi}\right)^{\frac{3}{2}} \label{m}\\
t&=&\frac{1}{4 \pi ^{3/2} s^{3/2}} \left(-\pi ^2 q^2 +\pi s + 3s^2 \right) \label{deft}\\
\Phi &=& q \sqrt{\frac{\pi}{s}} \label{defphi}\\
c_\Phi &=& 2s \frac{-q^2 \pi ^2 + \pi s +3s^2}{q^2 \pi ^2 - \pi s +3s^2} \label{defC}
\end{eqnarray} 

Before we proceed further, let us first put the analogy
between the variables corresponding
to RN-AdS black holes and fluid systems (that is 
encountered in standard thermodynamics )in a
tabular form:
\begin{table}[h]
\caption{Analogy between black hole parameters
and standard thermodynamic variables}
\centering                          
\begin{tabular}{|c|c|}            
\hline        
  RN-Ads Black holes &  pure fluids\\ [.05ex]
\hline                              
mass (m)        & Energy (E)\\
Temperature (t) & Temperature (T) \\ 
Entropy(s)      & Entropy (S) \\
Potential (-$ \Phi $)  & Pressure(P) \\
Charge (q) & Volume (V) \\ [.05ex]         
\hline                              
\end{tabular}
\end{table}\\
Now from (\ref{deft})  we observe that for nonextremality (which here implies $t>0$), we have
\begin{eqnarray}
&&\left(-\pi ^2 q^2 +\pi s + 3s^2 \right) > 0 \Rightarrow s > \frac{\pi}{6} \left(\sqrt{1+12q^2} - 1 \right)
\label{ext}
\end{eqnarray}
The critical points (points of phase transition) can be ontained by finding the roots of the denominator of the heat capacity and are given by 
\begin{equation}
s_\pm = \frac{\pi}{6}\left(1 \pm \sqrt{1-12q^2}\right)
\label{s+-}
\end{equation} 
For these roots to be a real quantity we must have $q\leq \frac{1}{2\sqrt{3}}$. We will denote the limiting value as the critical charge, 
\begin{equation}
q_{cri}= \frac{1}{2\sqrt{3}} = 0.289
\label{qc}
\end{equation}
One can easily show that for $q<q_{cri}$ both $s_\pm$ satisfy the nonextremality condition ({\ref{ext}}).

\section{Analysis of phase transition using Ehrenfest's equations }

Phase transition is one of the most interesting phenomena of a thermodynamic system. When we identify black holes as thermodynamic objects it is very natural to investigate whether they also undergo any kind of phase transition or not. A phase transition is denoted by a discontinuity of a state space variable, specially heat capacity. Expression (\ref{defC}) reveals that the heat capacity has poles given by (\ref{s+-}). In the physical domain $q<q_{cri}$ those poles are real. A phase transition is therefore expected on these general grounds.
\begin{figure}[h]
\centering
\includegraphics[scale=.75]{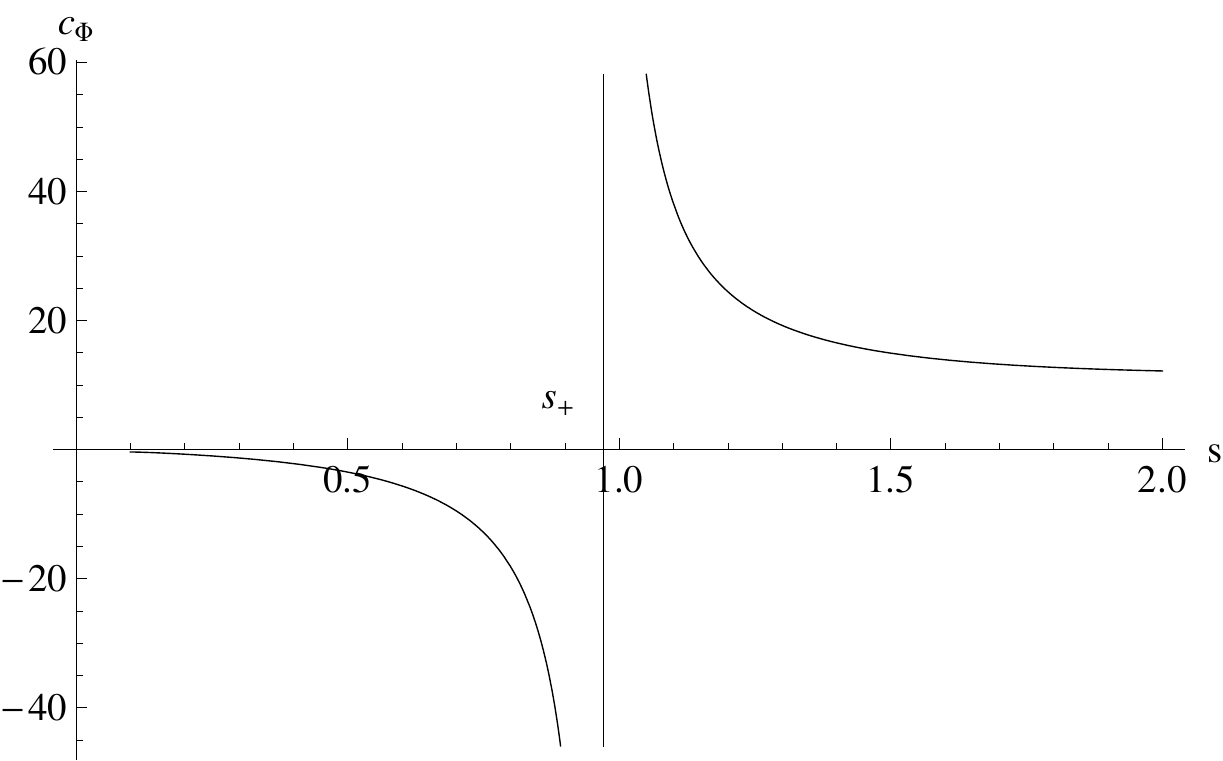}
\caption{variation of heat capacity ($c_\Phi $) with entropy ($s$) for $q=0.15$}
\label{cq}
\end{figure}

\begin{figure}[h]
\centering
\includegraphics[scale=1]{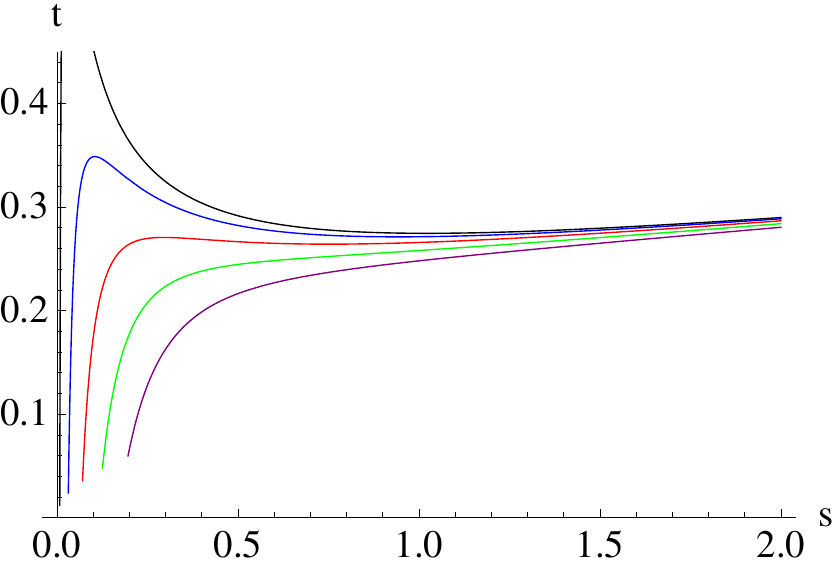}\\
\caption{Variation of temperature,  with entropy for different values of q; black (0.05), blue (0.10), red (0.15), green (0.20), purple (0.25)}
\label{ts}
\end{figure}

We will now study this phase transition using the approach of {\cite{Modak1,Modak2}}. The plot of heat capacity ($c_\phi$) against entropy ($s$) clearly illuminates a smeared discontinuity in $c_\Phi$ ($fig.$\ref{cq}) which  strongly indicates the onset of a phase transition. We will discuss this graph elaborately later on. For the present having indicated the possibility of a phase transition it is essential to study its nature- first order or second order. To see this let us first focus on the entropy temperature relationship ($fig.${\ref{ts}}). It is clear from the graph that there is no discontinuity, so that a first order transition is ruled out.\\ 

Next let us check whether it is a second order transition or not. This is done by adopting Ehrenfest's scheme \cite{Thu,zeman}. Before proceeding further we present a general discussion based on $fig.$\ref{cq}.The smeared discontinuity here is different from the true discontinuity that occurs in continuous phase transitions of classical type.  Such transitions occur
in standard fluid systems and are studied by using Ehrenfest's scheme. In the present context we thus have two options:
either to abandon this scheme or to take the Ehrenfest's scheme as an approximation by a suitable interpretation. We
adopt the second option. In considering Ehrenfest's scheme as an approximation we recall that something similar also occurs in glass where the discontinuities are smeared \cite{Thu}. In applying Ehrenfest's scheme we face two problems:
first, the discontinuities are smeared and secondly, $c_{\Phi}$ attains infinite values along the phase transition line. Both these issues are evaded by looking at those points on the curve which have (i) vanishingly small slope and (ii) are close to the critical point. This is similar in spirit to what is done for glassy phase transitions \cite{Thu}. The only difference is that for the glassy case, contrary to the present situation, $C_{P}$ is always finite \cite{Jack}. However since the divergence for $c_{\Phi}$ occurs in the smeared region which is anyway avoided (like in the case of glassy transitions\cite{Thu}), we feel that it should be inconsequential to our analysis. 
 
The first and second Ehrenfest's equations for RN-AdS black holes are
\begin{eqnarray}
&&-\left(\frac{\partial \Phi}{\partial t}\right)_s = \frac{1}{tq} \frac{c_{\Phi _2}-c_{\Phi _1}}{\beta _2-\beta _1} \label{eh1}\\
&&-\left(\frac{\partial \Phi}{\partial t}\right)_q = \frac{\beta _2-\beta _1}{\kappa _2 - \kappa _1}
\label{eh2}
\end{eqnarray}  
The subscript 1 and 2 represent phase 1 and 2 respectively. The new variables $\beta$ and $\kappa$ correspond volume expansivity and isothermal compressibility in statistical thermodynamics. They are given by (see Appendix A)
\begin{eqnarray}
\beta &= \frac{1}{q} \left( \frac{\partial q}{\partial t}\right)_\Phi &=\frac{4\pi ^{3/2} s^{3/2}}{\pi ^2 q^2 -\pi s + 3 s^2} \nonumber \\
\kappa &= -\frac{1}{q} \left(\frac{\partial q}{\partial \Phi} \right)_t &= \frac{\sqrt{s}}{q\sqrt{\pi}} \frac{3 \pi ^2 q^2 -\pi s + 3 s^2}{\pi ^2 q^2 -\pi s + 3 s^2}
\label{bk}
\end{eqnarray}

\begin{figure}[h]
\centering
\includegraphics[scale=.75]{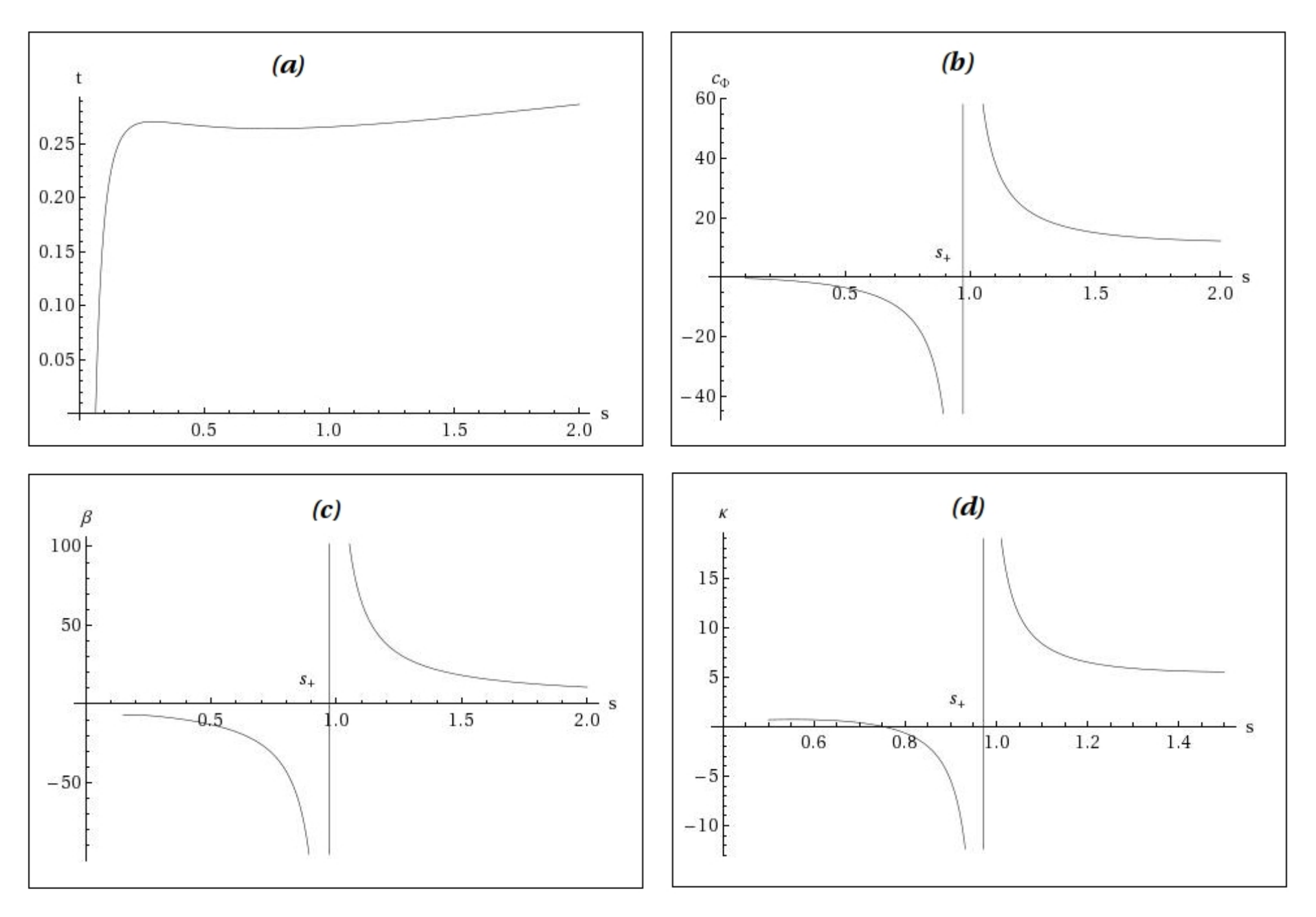}\\
\caption{Variation of (a) temperature, (b) heat capacity at constant potential, (c)volume expansivity and (d) isothermal compressibility with entropy for q = 0.15}
\label{graph}
\end{figure}

With the help of ({\ref{deft},\ref{defphi}}) and simple laws of differential calculas one can easily obtain the following expressions after a little mathematical manipulation (see Appendix B).
\begin{eqnarray}
&&\left(\frac{\partial \Phi}{\partial t}\right)_s = -\frac{2s}{q} \label{e1}\\
&&\left(\frac{\partial \Phi}{\partial t}\right)_q = -\frac{4 \pi ^2 qs}{3 \pi ^2 q^2 - \pi s +3s^2} \label{e2} 
\end{eqnarray}

\begin{figure}[h]
\centering
\includegraphics[scale=.8]{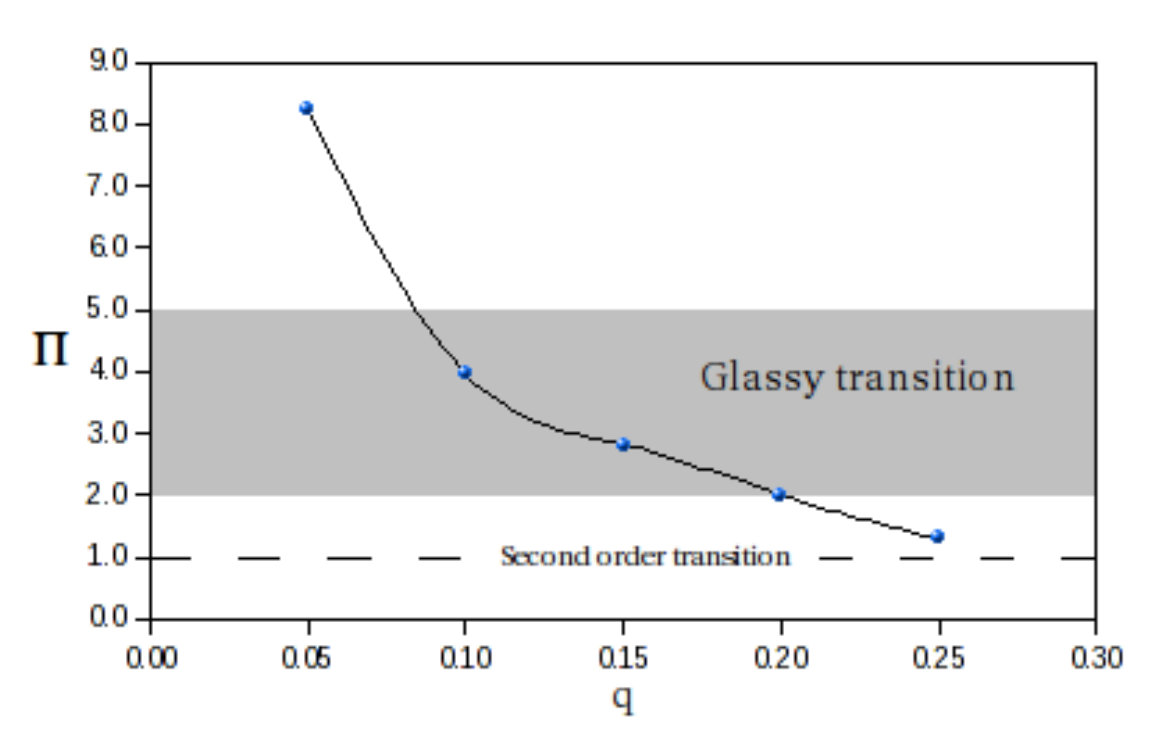}
\caption{change of PD ratio with q}
\label{PD}
\end{figure}

These expressions correspond to the LHS of the first and second Ehrenfest's equations (\ref{eh1},\ref{eh2}). It is now simple to show that the LHS of both (\ref{e1},\ref{e2}), evaluated at $s_+$, are identical (see Appendix C), i.e.
\begin{eqnarray}
\left[\left(\frac{\partial \Phi}{\partial t}\right)_q \right]_{s_+} = -\frac{4 \pi ^2 qs_+}{3 \pi ^2 q^2 - \pi s_+ +3s_+^2} = -\frac{2s_+}{q} = \left(\frac{\partial \Phi}{\partial t}\right)_{s_+} \label{eLHS}
\end{eqnarray}
Similarly one can obtain the RHS also in terms of $q$ and $s$. The RHS of Ehrenfest's equations contain entities with subscripts 1 and 2. To verify Ehrenfest's equations we take points as just described above. Furthermore, to check the robustness of the scheme we take, for each graph corresponding to a particular value of the charge q (within the specified range ($0<q<q_{cri},q_{cri} =0.289, ~see (\ref{qc})$) three sets or pair of points on the phase transition curve , which satisfy the two criteria of vanishingly small slope and closeness to the critical point. It is reassuring to note that almost identical results are obtained for these distinct sets of points as elaborated below.\\
Let us now consider the first Ehrenfest's equation (\ref{eh1}). The results are tabulated below in $table$(\ref{E1}), $table$(\ref{E2}) and $table$(\ref{E3}) corresponding to the three pairs of points on the phase transition curve. Moreover, five distinct values of q within the physical range ($0<q<q_{cri},q_{cri} =0.289, ~see (\ref{qc})$) are chosen.
\begin{subtables}
\begin{table}[htb]
\caption{First Ehrenfest's equation}   
\centering                          
\begin{tabular}{c c c c c c c}            
\hline\hline                        
q & $s_+$(\ref{s+-})  & t(\ref{deft}) &($c_{\Phi _2}-c_{\Phi _1}$) & ($\beta _2 - \beta _1 $) & LHS ($\frac{2s_+}{q}$) & RHS \\ [0.05ex]
\hline                              
0.05 & 1.0393 & 0.2746 & (12.7204-(-0.1120)) & (18.1917-(-4.1191)) & 41.5714 & 41.8880 \\
0.10 & 1.0148 & 0.2714 & (12.5076-(-0.1295)) & (17.6391-(-4.8635)) & 20.2956 & 20.6949 \\ 
0.15 & 0.9710 & 0.2654 & (12.1647-(-0.3933)) & (16.7890-(-6.9595)) & 12.9462 & 13.2807 \\
0.20 & 0.9012 & 0.2557 & (11.7078-(-1.2456)) & (15.7279-(-13.7074)) & 9.0117 &  8.6043 \\
0.25 & 0.7854 & 0.2387 & (11.1571-(-5.1659)) & (12.5108-(-35.5059)) & 6.2832 &  5.6958 \\ [0.05ex]         
\hline                              
\end{tabular}
\label{E1}          
\end{table}
\begin{table}[htb]
\caption{First Ehrenfest's equation}   
\centering                          
\begin{tabular}{c c c c c c c}            
\hline\hline                        
q & $s_+$(\ref{s+-})  & t(\ref{deft}) &($c_{\Phi _2}-c_{\Phi _1}$) & ($\beta _2 - \beta _1 $) & LHS ($\frac{2s_+}{q}$) & RHS \\ [0.05ex]
\hline                              
0.05 & 1.0393 & 0.2746 & (12.5962-(-0.1120)) & (16.8319-(-4.1191)) & 41.5714 & 44.1750 \\
0.10 & 1.0148 & 0.2714 & (12.3990-(-0.1295)) & (16.3792-(-4.8635)) & 20.2956 & 21.7338 \\ 
0.15 & 0.9710 & 0.2654 & (12.0804-(-0.3933)) & (15.6765-(-6.9595)) & 12.9462 & 13.8400 \\
0.20 & 0.9012 & 0.2557 & (11.6545-(-1.2456)) & (14.7882-(-13.7074)) & 9.0117 &  8.8514 \\
0.25 & 0.7854 & 0.2387 & (11.1388-(-5.1659)) & (11.9732-(-35.5059)) & 6.2832 &  5.7539 \\ [0.05ex]         
\hline                              
\end{tabular}
\label{E2}          
\end{table}
\begin{table}[htb]
\caption{First Ehrenfest's equation}   
\centering                          
\begin{tabular}{c c c c c c c}            
\hline\hline                        
q & $s_+$(\ref{s+-})  & t(\ref{deft}) &($c_{\Phi _2}-c_{\Phi _1}$) & ($\beta _2 - \beta _1 $) & LHS ($\frac{2s_+}{q}$) & RHS \\ [0.05ex]
\hline                              
0.05 & 1.0393 & 0.2746 & (12.8692-(-0.0536 )) & (18.1917-(-3.7638)) & 41.5714 & 42.8658 \\
0.10 & 1.0148 & 0.2714 & (12.6385-(-0.0415 )) & (17.6391-(-4.5394)) & 20.2956 & 21.0686 \\ 
0.15 & 0.9710 & 0.2654 & (12.2677-(-0.4394 )) & (16.7890-(-6.9140)) & 12.9462 & 13.4643 \\
0.20 & 0.9012 & 0.2557 & (11.7758-(-1.3032 )) & (15.7279-(-14.8388)) & 9.0117 &  8.3662 \\
0.25 & 0.7854 & 0.2387 & (11.1856-( -5.3691)) & (13.1106-(-34.9738)) & 6.2832 &  5.7685 \\ [0.05ex]         
\hline                              
\end{tabular}
\label{E3}          
\end{table}
\end{subtables}
Our results show that there is reasonably good agreement between the LHS and RHS of (\ref{eh1}) for all cases. This shows the validity of the first Ehrenfest equation as well as the robustness of the approach.\\
We now consider the second Ehrenfest relation (\ref{eh2}). Here we find that there is a mismatch between the LHS and RHS. This mismatch is measured by a new parameter called the  Prigogine-Defay (PD) ratio,defined as \cite{Gupta}  
\begin{equation}
\Pi = \frac{\Delta c_\Phi \Delta \kappa}{t q \left(\Delta \beta\right)^2}
\label{pd}
\end{equation}
Our findings are enumerated in $table$(\ref{F1}), $table$(\ref{F2}) and $table$(\ref{F3})
\begin{subtables}
\begin{table}[htb]
\caption{Second Ehrenfest's equation and PD ratio ($\Pi $)}   
\centering                          
\begin{tabular}{c c c c c c}            
\hline\hline                        
q & LHS & $\Delta \kappa (\kappa_2-\kappa_1)$ & RHS (\ref{eh2}) $\left(\frac{\Delta \beta}{\Delta \kappa}\right)$ & RHS(\ref{eh1}) $ \frac{1}{tq} \frac{\Delta c_\Phi}{\Delta \beta} $ & $\Pi$ \\ [0.05ex]
\hline                              
0.05 & 41.5714 & (13.7885-9.3295)   & 5.0036 & 41.8880 & 8.3716 \\
0.10 & 20.2956 & (7.5093-3.1940)    & 5.2145 & 20.6949 & 3.9687 \\ 
0.15 & 12.9462 & (5.6105-0.5946)    & 4.7347 & 13.2807 & 2.8049 \\
0.20 & 9.0117  & (4.7422-(-2.0212)) & 4.3522 & 8.6043  & 1.9770 \\
0.25 & 6.2832  & (4.2400-(-8.4715)) & 3.7774 & 5.6958  & 1.5078 \\ [0.05ex]         
\hline                              
\end{tabular}
\label{F1}          
\end{table}
\begin{table}[htb]
\caption{Second Ehrenfest's equation and PD ratio ($\Pi $)}   
\centering                          
\begin{tabular}{c c c c c c}            
\hline\hline                        
q & LHS & $\Delta \kappa (\kappa_2-\kappa_1)$ & RHS (\ref{eh2}) $\left(\frac{\Delta \beta}{\Delta \kappa}\right)$ & RHS(\ref{eh1}) $ \frac{1}{tq} \frac{\Delta c_\Phi}{\Delta \beta} $ & $\Pi$ \\ [0.05ex]
\hline                              
0.05 & 41.5714 & (13.9648-9.3295)    & 4.5199 & 44.1750 & 9.7735 \\
0.10 & 20.2956 & (7.5183-3.1940)     & 4.9124 & 21.7338 & 4.4243 \\ 
0.15 & 12.9462 & (5.5480-0.5946)     & 4.5698 & 13.8440 & 3.0286 \\
0.20 & 9.0117  & (4.6461-(-2.0212))  & 4.2740 & 8.8514  & 2.0710 \\
0.25 & 6.2832  & (4.1327-(-8.4715))  & 3.7670 & 5.7539  & 1.5275 \\ [0.05ex]         
\hline                              
\end{tabular}
\label{F2}          
\end{table}
\begin{table}[htb]
\caption{Second Ehrenfest's equation and PD ratio ($\Pi $)}   
\centering                          
\begin{tabular}{c c c c c c}            
\hline\hline                        
q & LHS & $\Delta \kappa (\kappa_2-\kappa_1)$ & RHS (\ref{eh2}) $\left(\frac{\Delta \beta}{\Delta \kappa}\right)$ & RHS(\ref{eh1}) $ \frac{1}{tq} \frac{\Delta c_\Phi}{\Delta \beta} $ & $\Pi$ \\ [0.05ex]
\hline                              
0.05 & 41.5714 & (13.6277-9.3362)    & 5.1159  & 42.8658  &  8.3789 \\
0.10 & 20.2956 & (7.5318-3.1314)     & 5.0400  & 21.0686  &  4.1802 \\ 
0.15 & 12.9462 & (5.7102-0.5086)     & 4.5569  & 13.4643  & 2.9547 \\
0.20 & 9.0117  & (4.8740-(-2.2071))  & 4.3167  &  8.3662  &  1.9381  \\
0.25 & 6.2832  & (4.3772-(-8.9680))  & 3.6031  &  5.7685  &  1.6010 \\ [0.05ex]         
\hline                              
\end{tabular}
\label{F3}          
\end{table}
\end{subtables} 
For second order phase transition the PD ratio is 1. For a PD ratio 2-5 the phase transition is called glassy \cite{Jack}. We considered five trial values for q and obtained different values for the PD ratio. The phase transition is glassy for a particular range of q. The PD ratio decreases with q and the limiting value (for $q=q_{cri}$) is $>1$, i.e. the phase transition is never second order (see $fig$\ref{PD},which corresponds to the values given in table 3a). Once again we find self consistent results for all pairs of points. The phase transition is glassy within the appropriate range($0.1\leq q \leq 0.2$). 

\section{Study of thermodynamics using state space geometry}
An alternative way to study thermodynamics is using the state space geometry \cite{Rupp}-\cite{Rupp2}. In this section we will study the nature of the corresponding scalar curvature and interpret the obtained result. The metric as defined by Ruppeiner \cite{Rupp,Rupp2} is given by,
\begin{equation}
g_{ij} = -\frac{\partial ^2 s(x)}{\partial x^i \partial x^j}
\label{gR}
\end{equation}
where $x^i$-s are the extensive variables of the system. It is convenient to work with the Weinhold metric ($g^W_{ij}$) \cite{Wein} defined as
\begin{equation}
g^W_{ij} = \frac{\partial ^2 m(x)}{\partial x^i \partial x^j}
\label{Wg}
\end{equation}
The line elements in Ruppeiner geometry and the Weinhold geometry are conformally related by \cite{Mru,PS}
\begin{equation}
dS_R^2 = \frac{1}{t} dS_W^2
\label{RW}
\end{equation}
To evaluate the Weinhold metric we will consider $x^1 = s$ and $x^2 = q$. One can easily evaluate the Weinhold metric using ({\ref{m}}). Divide this by $t$ ({\ref{deft}}) to get the Ruppeiner metric
\begin{eqnarray}
&& g_{ss} = \frac{1}{2s} \frac{\left(3\pi ^2 q^2 - \pi s +3s^2 \right)}{\left(-\pi ^2 q^2 + \pi s +3s^2\right)} \nonumber \\
&& g_{sq} = - \frac{2\pi ^2 q}{\left(-\pi ^2 q^2 + \pi s +3s^2\right)} = g_{qs}\\
&& g_{qq} = \frac{4\pi ^2 s}{\left(-\pi ^2 q^2 + \pi s +3s^2\right)} \nonumber
\label{gsq}
\end{eqnarray}
Observe that all the metric components have an identical denominator which is, furthermore, same as the denominator appearing in the heat capacity ({\ref{defC}}). The determinant of the above metric is given by
\begin{equation}
g=2\pi ^2 \frac{\left(\pi ^2 q^2 - \pi s +3s^2\right)}{\left(-\pi ^2 q^2 + \pi s +3s^2\right)^2}
\label{g}
\end{equation}

\begin{figure} [h]
\centering
\includegraphics[scale=1]{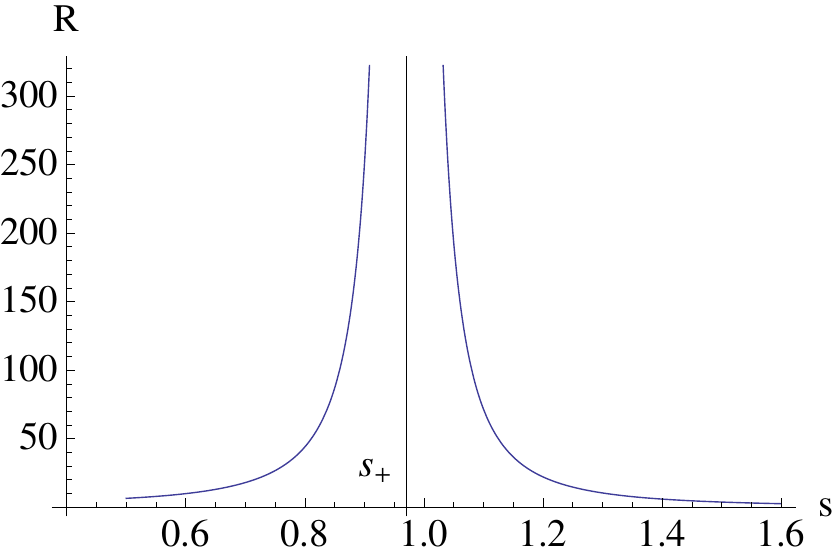}
\caption{Ruppeiner curvature as a function of entropy. The divergence is exactly at the same point ($s_+$) where $c_\Phi$ diverges.}
\label{RC}
\end{figure}

The resulting scalar curvature in terms of $q$ and $s$ follows after some algebra,
\begin{equation}
R=-9 s \frac{ \left( \pi ^2 q^2 -\pi s + s^2 \right) \left( \pi ^2 q^2+3s^2\right) }{ \left(-\pi ^2 q^2 + \pi s +3s^2 \right) \left(\pi ^2 q^2 - \pi s +3s^2 \right)^2}
\label{R}
\end{equation} 
Here we have set the Boltzman constant ($k_B$) to 1. If it is set equals to $\frac{1}{\pi}$ the expression for the scalar curvature reproduces that obtained in \cite{Aman}. The first part of this denominator also appears in the expression of the temperature ({\ref{deft}}) but we cannot put it equal to zero due to the nonextremal condition (\ref{ext}). The second part is identical with the denominator of the heat capacity ({\ref{defC}}). Hence the curvature will diverge exactly at those points at which the heat capacity diverges ($fig.$\ref{RC}).

\section{Discussion}
We have obtained different thermodynamic entities like temperature, potential and heat capacity at constant potential for a RN AdS black hole from the first law of black hole thermodynamics. The heat capacity shows a smeared discontinuity at some values of the entropy ({\ref{s+-}}) which we interpret as a signature of phase transition. Since this critical entropy cannot be a complex quantity, it also predicts an upper bound for the charge ({\ref{qc}}) beyond which the analysis becomes unphysical and hence is ignored. From the smoothness of entropy temperature curve ($fig.$ {\ref{ts}}) one can easily guess that the phase transition is not first order. We further study the type of this phase transition using Ehrenfest's equations. The first Ehrenfest's equation ({\ref{E1}}) is well satisfied. The LHS and RHS of the second Ehrenfest equation do not match so finely. Consequently a standard second order phase transition is also ruled out.

However we recall that there is another type, the so called glassy phase transition \cite{Jack}. This is a dynamical freezing transition that occures when a liquid is supercooled. It behaves as a smeared second order phase transition where the first Ehrenfest's equation is always satisfied while the second is not. The Prigogine-Defay (PD) ratio that measures the deviation from the second Ehrenfest's equation lies in the range 2 to 5 \cite{Th,Thu}. We consider different trial values for the charge and obtain different PD ratios. The ratio decreases with increasing charge ($fig${\ref{PD}}) eventually termainating at a value $>1$. The values of the ratio show that for an intermediate range of charge the transition is glassy (PD ratio 2-5) but it is never a second order transition. 

While applying Ehrenfest's scheme for glassy type transitions as found here, one must be careful about choosing the values of heat capacity, volume expansivity or isothermal compressibility for the two phases since these variables make smeared jumps. Inspired by the approach in glassy transitions \cite{Th,Thu},we have attempted to interpret the Ehrenfest scheme in the present context. We avoid the smeared zone and, for numerical computation, consider those sets of points that have a vanishingly small slope on the phase transition curve and are close to the critical point. To check the robustness of our approximation, we have taken three distinct pairs of points satisfying this criteria. For all five values of the charge (q) in the allowed range we obtain self consistent results for both Ehrenfest's relations. Our conclusion is that, while the first relation (\ref{eh1}) always holds, the second one (\ref{eh2}) is violated. The violation is such that the phase transition is glassy with the allowed PD ratio(2-5) for values of q satisfying $0.1\leq q \leq 0.2$. The phase transition is neither first order nor second order.  

We have also studied aspects of phase transition using methods of state  space geometry developed in \cite{Rupp}. We find that both the scalar curvature and the heat capacity have a common denominator and hence diverge at identical points. This shows that a divergence in the scalar curvature corresponds to a discontinuity in the heat capacity thereby suggesting the occurrence of a phase transition. Furthermore there is another term in the denominator of the scalar curvature which one might expect to be a possible source of divergence, but a careful inspection shows that the same expression arises in the expression for the temperature ({\ref{deft}}) and hence according to the third law of thermodynamics we cannot put it equal to zero. Thus from the nature of Ruppeiner curvature we can easily distinguish the critical points and get information about the occurance of phase transition.
\section*{Appendix A}
Here we provide a short derivation for the volume expansivity ($\beta$) and isothermal compressibility ($\kappa$) appearing in (\ref{bk}).

We have the expression for $t$ in terms of $s$ and $q$ ({\ref{deft}}). With the help of ({\ref{defphi}}) we can write $t$ as a function of $\Phi$ and  $q$.
\begin{equation}
t(\Phi ,q) = \frac{1}{4\pi}\left( \frac{\Phi}{q} - \frac{\Phi ^3}{q^2}+ 3 \frac{q}{\Phi} \right) \label{tpq} 
\end{equation} 
Taking a partial derivative and using ({\ref{defphi}}), we get
\begin{equation}
\left( \frac{\partial t}{\partial q}\right)_\Phi = \frac{1}{4\pi}\left(-\frac{1}{q}\sqrt{\frac{\pi}{s}} + q \left(\frac{\pi}{s}\right)^{3/2}+\frac{3}{q}\sqrt{\frac{s}{\pi}} \right) \label{a1}
\end{equation}
which directly gives
\begin{equation}
\beta = \frac{1}{q} \left( \frac{\partial q}{\partial t}\right)_\Phi = \frac{4\pi ^3/2 s^3/2}{\pi ^2 q^2 -\pi s + 3 s^2} \label{beta}
\end{equation}

One can obtain an expression for $\kappa$ also in a similar way. To get $\kappa$ we will use a well known formula
\begin{eqnarray}
\left(\frac{\partial t}{\partial \Phi} \right)_q \left(\frac{\partial \Phi}{\partial q} \right)_t \left(\frac{\partial q}{\partial t} \right)_\Phi &=& -1 \nonumber \\
\Rightarrow \left(\frac{\partial \Phi}{\partial q} \right)_t &=& -\frac{\left(\frac{\partial t}{\partial q} \right)_\Phi}{\left(\frac{\partial t}{\partial \Phi} \right)_q} \label{cr}
\end{eqnarray}
This relationship enables us to evaluate $\left(\frac{\partial \Phi}{\partial q} \right)_t$ in terms of $\left(\frac{\partial q}{\partial t} \right)_\Phi$ and $\left(\frac{\partial t}{\partial \Phi} \right)_q$. The first one is already derived ({\ref{a1}}). The other term is given by (using ({\ref{tpq}}) and ({\ref{defphi}}))
\begin{equation}
\left(\frac{\partial t}{\partial \Phi} \right)_q = \frac{1}{4 \pi} \left(\frac{1}{q} - \frac{3\pi q}{s} -\frac{3s}{\pi q}\right) \label{a2}
\end{equation} 
Combining ({\ref{cr}}),({\ref{a1}}) and ({\ref{a2}}) we finally obtain
\begin{equation}
\kappa = - \frac{1}{q} \left(\frac{\partial q}{\partial \Phi} \right)_t = \frac{\sqrt{s}}{q\sqrt{\pi}} \frac{3 \pi ^2 q^2 -\pi s + 3 s^2}{\pi ^2 q^2 -\pi s + 3 s^2}
\end{equation} 
\section*{Appendix B}
Here a brief derivation of (\ref{e1}) and ({\ref{e2}) is given.

Using ({\ref{deft}}) and ({\ref{defphi}}) we can easily express $t$ as a function of $\Phi$ and $s$ or $\Phi$ and $q$.
\begin{eqnarray}
t(\Phi ,s) &=& \frac{1}{4 \pi ^{3/2} s^{3/2}} \left(-\pi \Phi ^2 s +\pi s + 3s^2 \right) \\
t(\Phi ,q) &=& \frac{1}{4 \pi} \left(-\frac{\Phi ^3}{q}+\frac{\Phi}{q}+\frac{3q}{\Phi} \right)
\end{eqnarray}
From the above two expressions one can easily obtain 
\begin{eqnarray}
&\left(\frac{\partial t}{\partial \Phi}\right)_s &= -\frac{1}{2s}\Phi \sqrt{\frac{s}{\pi}} = -\frac{q}{2s} \label{b1} \\
&\left(\frac{\partial t}{\partial \Phi}\right)_q &= \frac{1}{4\pi}\left( \frac{1}{q} - \frac{3 \pi q}{s} - \frac{3s}{\pi q} \right) =- \frac{3\pi ^2 q^2 - \pi s + 3 s^2}{4 \pi ^2 qs} \label{b2}
\end{eqnarray}
which yield (\ref{e1}) and (\ref{e2}), respectively, on inversion.
\section*{Appendix C}
Here we show the equivalence of the LHS of the Ehrenfest's equations (\ref{eh1}) and (\ref{eh2}) evaluated at $s_+$.

By definition the heat capacity diverges at $s=s_+$. Hence from ({\ref{defC}}) we can write  
\begin{eqnarray}
q^2 \pi ^2 - \pi s_+ +3s_+^2 = 0 \Rightarrow \pi s_+ - 3s_+^2 = q^2 \pi ^2 
\end{eqnarray}
When we put this value of $\pi s_+ - 3s_+^2$ in the RHS of ({\ref{e2}}), it readily gives
\begin{eqnarray}
-\frac{4 \pi ^2 qs_+}{3 \pi ^2 q^2 - \pi s_+ +3s_+^2} = -\frac{4 \pi ^2 qs_+}{2 \pi ^2 q^2} = -\frac{2s_+}{q}
\end{eqnarray}
leading to ({\ref{eLHS}).

\section*{Acknowledgement}
 S. Ghosh and D. Roychowdhury like to thank Council for Scientific and Industrial Research (C.S.I.R.), India for financial support.  
 
\end{document}